\documentclass[aps,prb,superscriptaddress,amsmath,amssymb,twocolumn,a4paper,floatfix]{revtex4-1}
\usepackage{graphicx}
\usepackage{float}
\usepackage{hyperref}
\usepackage{bm}
\usepackage{tabularx}
\usepackage{modiagram}
\usepackage{ifthen}
\usepackage{multirow}
\usepackage{xcolor}
\usepackage{bbm}
\usepackage{accents}

\newcolumntype{Z}{>{\centering\let\newline\\\arraybackslash\hspace{0pt}}X}

\newcommand{\LocalState}[6]{
  \left|
  \begin{tikzpicture}[baseline=(a.center)]
    \draw[-] (0.3,0.5) -- (0.7,0.5);
    \draw[-] (0.0,0) -- (0.4,0);
    \draw[-] (0.6,0) -- (1.0,0);
    \ifthenelse{#1=1}{\filldraw (0.40, 0.5) circle (0.06);}{}
    \ifthenelse{#3=1}{\filldraw (0.10, 0.0) circle (0.06);}{}
    \ifthenelse{#5=1}{\filldraw (0.70, 0.0) circle (0.06);}{}
    \ifthenelse{#2=1}{\draw (0.60, 0.5) circle (0.06);}{}
    \ifthenelse{#4=1}{\draw (0.30, 0.0) circle (0.06);}{}
    \ifthenelse{#6=1}{\draw (0.90, 0.0) circle (0.06);}{}
    \node at (0.4,0.20) (a){};
  \end{tikzpicture}
  \right\rangle
}

\begin{document}
\title{Alleviating the sign problem in quantum Monte Carlo simulations
of spin-orbit-coupled multiorbital Hubbard models}
\author{Aaram J. Kim}
\email{aaram.kim@kcl.ac.uk}
\affiliation{Department of Physics, King's College London, Strand, London WC2R 2LS, UK}

\author{Philipp Werner} 
\affiliation{Department of Physics, University of Fribourg, Chemin du Mus\'ee 3, 1700 Fribourg, Switzerland}

\author{Roser Valent\'i}
\affiliation{Institut f\"ur Theoretische Physik, Goethe-Universit\"at Frankfurt, Max-von-Laue-Str. 1, 60438 Frankfurt am Main, Germany}

\begin{abstract}
	We present a strategy to alleviate the sign problem
  in continuous-time quantum Monte Carlo
(CTQMC) simulations of the dynamical-mean-field-theory (DMFT) equations
for the spin-orbit-coupled multi-orbital Hubbard model.  
We first identify the combinations of rotationally invariant Hund coupling terms present in the relativistic basis
which lead to a severe sign problem.  
Exploiting the fact that the average sign in CTQMC depends on the
choice of single-particle basis, we  propose a {\it
bonding-antibonding basis} $V_{j3/2\mathrm{BA}}$ which shows an improved
average sign compared to the widely used relativistic basis for most parameter
sets investigated.  We then generalize this procedure
by introducing a stochastic optimization algorithm
that exploits the space of single-particle bases
 and show that $V_{j3/2\mathrm{BA}}$ is
very close to optimal within the  parameter space investigated. 
Our findings enable more efficient DMFT simulations of materials with strong spin-orbit coupling.  
\end{abstract}

\maketitle

\section{Introduction}
Spin-orbit coupling (SOC) is an essential ingredient in the study of exotic
phases in correlated electron systems,\cite{WitczakKrempa:2014hz}
 such as unconventional
superconductivity in $4d$ transition-metal oxides,\cite{Mackenzie:2003er,Wang:2011cm,Watanabe:2013ge,Kim:2014hn,Yang:2014bc,Meng:2014jo,Kim:2015kf,Borisenko:2015ex}
topological phases of matter in quantum spin-Hall insulators,\cite{Kane:2005hl,Hasan:2010ku,Qi:2011hb}
excitonic
insulators,~\cite{Khaliullin:2013du,Kunes:2014ea,Sato:2015hq,Kim:2017kt,Sato:2019by} and
 Kitaev-model-based insulators,\cite{Jackeli:2009hz,rau2016spin,savary2016quantum,winter2017models,hermanns2018physics,takagi2019concept}
to mention a few.
A prototypical minimal
 model that includes the interplay between spin-orbit coupling
and correlations is the relativistic  multiorbital Hubbard model.
Its nonrelativistic counterpart has been intensively investigated in the past and shows a rich phase
diagram.\cite{Werner2008,Medici2011,Kunes:2014ea,Hoshino2016}
The relativistic multiorbital Hubbard model,\cite{Watanabe:2013ge,Meng:2014jo,Sato:2015hq,Shinaoka:2015cg,Kim:2017kt,Sato:2019by} however, is much less understood since 
the choice of algorithms is strongly limited 
due to the extra computational complications associated with (multiple-)spin-orbit-coupled
 degrees of freedom.

One of the promising formalisms to investigate the
Hubbard model and its generalizations is the dynamical 
mean-field theory (DMFT)~\cite{Georges:1996zz,Kotliar2004} that has  
provided important insights into multiorbital physics
also in combination with
{\it ab initio} calculations for real materials.\cite{Kotliar:2006fl,yin2011kinetic,lechermann2016electron}
The continuous-time quantum Monte Carlo method (CTQMC),\cite{Gull:2011jd} 
particularly the hybridization expansion algorithm (CTHYB),\cite{Werner:2006ko,Werner:2006iz} is the most widely used impurity solver in multiorbital DMFT calculations. 
However, CTHYB suffers
 from the notorious sign problem when the SOC is included in the calculations.
The sign problem 
grows exponentially with inverse
temperature~\cite{Loh:1990hj,Troyer:2005hv} 
and 
typically prevents the study of low-temperature 
symmetry-broken phases.
Alleviating the sign problem in CTHYB would 
help improve our 
understanding of phenomena determined
by the interplay of spin-orbit coupling and correlations.\cite{Mackenzie:2003er,Kim:2015kf,Borisenko:2015ex,2015PhRvL.115x7001H,Kim:2018hj}

For quantum Monte Carlo (QMC) algorithms based on auxiliary fields, 
there have been various successful advances which
unveiled the origin of the sign problem and suggested a solution in specific
cases,\cite{Chandrasekharan:ul,Wang:2015hm,Iglovikov:2015bj,Kung:2016he,Hann:2017ca,Yoo:2005ib,Alet:2016dd,Huffman:2016ih,Honecker:2016bs}
including the recently developed idea of Majorana
symmetry.\cite{Li:2015jf,Li:2016fi,Wei:2016cb}  The rotationally invariant
Hund coupling in the SO-coupled multiorbital Hubbard system,
 however, generates rather complex interaction terms.
Furthermore, the nonlocal-in-time expansion scheme of CTHYB makes it difficult to track the origin of the fermionic sign on the world line configuration. 

In this paper, we systematically study the nature of the sign problem of the
CTHYB for the SO-coupled three-orbital Hubbard model and propose
a strategy to alleviate it.
We employ a numerical sign-optimization scheme, called spontaneous
perturbation stochastic approximation (SPSA),\cite{Spall:1992fh} to determine the optimal basis 
in terms of average sign. Remarkably, this optimal basis can be well approximated by a simple one,
 which we denote as $V_{j3/2\mathrm{BA}}$. 
The $V_{j3/2\mathrm{BA}}$ basis is obtained from the relativistic basis by a bonding-antibonding transformation. 

\section{Model and Method}
Our model Hamiltonian is composed of the three terms $\mathcal{H} = \mathcal{H}_{\text{t}} + \mathcal{H}_{\text{soc}} + \mathcal{H}_{\text{int}}$, where $\mathcal{H}_{\text{t}}$, $\mathcal{H}_{\text{soc}}$, and $\mathcal{H}_{\text{int}}$ are, respectively,
the electron hopping, SOC, and Coulomb interaction terms.
For the noninteracting electron hopping part we assume a degenerate semicircular density of states $\rho(\omega) = (1/\pi D)\sqrt{1-(\omega/D)^2}$ and set the half-bandwidth $D$ as the unit of energy.
In the orbital-spin basis $V_{\mathrm{os}}$, $\mathcal{H}_{\mathrm{soc}}$ has the form 
\begin{equation}
	\mathcal{H}_{\text{soc}} = -\lambda\sum^{}_{i}\sum^{}_{\substack{\alpha\alpha'\\ \sigma\sigma'}}c^{\dagger}_{i\alpha\sigma}\langle \alpha\sigma|\mathbf{L}_{\text{eff}}\cdot \mathbf{S}|\alpha'\sigma'\rangle c^{}_{i\alpha'\sigma'}~,
	\label{}
\end{equation}
where $\mathbf{L}_{\text{eff}}$ is the $l=1$ orbital angular momentum operator and $\mathbf{S}$ is the spin operator.
$c^{}_{i\alpha\sigma}$ ($c^{\dagger}_{i\alpha\sigma}$) is the electron annihilation (creation) operator of orbital $\alpha$ ($yz,zx,xy$) and spin $\sigma$ ($\uparrow,\downarrow$) at lattice site $i$. 
We introduce the Slater-Kanamori form of the Coulomb interaction including the spin-flip and pair-hopping terms: 
\begin{eqnarray}
	\mathcal{H}_{\text{int}} &=& U\sum^{}_{i,\alpha}n_{i\alpha\uparrow}n_{i\alpha\downarrow} + \sum^{}_{\substack{i,\alpha<\alpha' \\ \sigma\sigma'}}(U'-J_{\rm H}\delta_{\sigma\sigma'})n_{i\alpha\sigma}n_{i\alpha'\sigma'} \nonumber\\
	&& -J_{\rm H}\sum^{}_{i,\alpha<\alpha'}(c^{\dagger}_{i\alpha\uparrow}c^{\dagger}_{i\alpha'\downarrow}c^{}_{i\alpha'\uparrow}c^{}_{i\alpha\downarrow} + \text{H.c.})\nonumber\\
	&& +J_{\rm H}\sum^{}_{i,\alpha<\alpha'}(c^{\dagger}_{i\alpha\uparrow}c^{\dagger}_{i\alpha\downarrow}c^{}_{i\alpha'\downarrow}c^{}_{i\alpha'\uparrow} + \text{H.c.})~.
  \label{eqn:kanamori}
\end{eqnarray}
Here, $U$ ($U'$) is the
intraorbital (interorbital) Coulomb repulsion, and $J_{\text{H}}$ is the strength of the Hund coupling.
To make $\mathcal{H}_{\mathrm{int}}$ rotationally invariant, we choose $U'=U-2J_{\mathrm{H}}$~.
This model Hamiltonian covers a wide range of $4d$ and $5d$ materials listed in Appendix~\ref{app:material} by tuning the electron density, $\lambda$, $U$, and $J_{\mathrm{H}}$~.

We solve the model Hamiltonian within the framework of DMFT.\cite{Georges1996}
The lattice model is mapped onto a quantum impurity model 
by means of a self-consistency relation.
The effective action of the quantum impurity model is written in terms of the local action $\mathcal{S}_{\mathrm{loc}}$ and the hybridization function $\Delta$,
\begin{eqnarray}
	\mathcal{S}_{\text{imp}} &=& \mathcal{S}_{\text{loc}} \nonumber\\
	&+& \sum^{}_{\substack{\alpha\alpha'\\\sigma\sigma'}}\int_{0}^{\beta}d\tau d\tau'~c^{\dagger}_{\alpha\sigma}(\tau)\Delta_{\alpha\sigma,\alpha'\sigma'}(\tau-\tau')c^{}_{\alpha'\sigma'}(\tau')~.\nonumber\\
	\label{}
\end{eqnarray}
Here, $c^{}_{\alpha\sigma}$ and $c^{\dagger}_{\alpha\sigma}$ are the fermionic 
annihilation and creation operators 
for orbital $\alpha$ and spin $\sigma$ for the impurity site.
CTHYB solves this impurity problem by performing an expansion of the partition function in powers of the hybridization function,
\begin{eqnarray}
	\mathcal{Z}_{\text{imp}} &=& \mathcal{Z}_{\mathrm{bath}}\sum^{\infty}_{k=0}\frac{1}{(k!)^2}\sum^{}_{\substack{\alpha_1\alpha'_1\\\sigma_1\sigma'_1}}\cdots\sum^{}_{\substack{\alpha_k\alpha'_k\\\sigma_k\sigma'_k}} \nonumber\\
	&\times&\int_{0}^{\beta}d\tau_1\int_{0}^{\beta}d\tau'_1\cdots \int_{0}^{\beta}d\tau_k\int_{0}^{\beta}d\tau'_k~\det\left[ \underline{\Delta}^{(k)} \right]\nonumber\\
	&\times& \left\langle\mathcal{T}_\tau c^{}_{\alpha'_1\sigma'_1}(\tau'_1)c^{\dagger}_{\alpha_1\sigma_1}(\tau_1) \cdots c^{}_{\alpha'_k\sigma'_k}(\tau'_k)c^{\dagger}_{\alpha_k\sigma_k}(\tau_k)\right\rangle_{\mathcal{S}_{\mathrm{loc}}}.\nonumber\\
	\label{eqn:Zimp}
\end{eqnarray}
Here, $\underline{\Delta}^{(k)}$ represents the $k\times k$ matrix whose elements are 
\begin{equation}
	\underline{\Delta}^{(k)}_{ij} = \Delta_{\alpha_i\sigma_i,\alpha'_j\sigma'_j}(\tau_i-\tau'_j)~.
	\label{}
\end{equation}
The hybridization function contains the relevant information about the continuous bath degrees of freedom. Since $\Delta$ is directly updated in the DMFT self-consistency loop, CTHYB does not rely on a discretization or truncation of the bath, in contrast to other impurity solvers like the exact diagonalization or the numerical renormalization group.\cite{Georges:1996zz,Gull:2011jd}
In that sense, CTHYB is numerically exact in treating the bath degrees of freedom.

The resulting partition function, Eq.~(\ref{eqn:Zimp}), can be expressed as a sum of configuration weights,  
\begin{equation}
	\mathcal{Z}_{\text{imp}} = \sum^{}_{\mathbf{x}} \omega(\mathbf{x})
	\label{}
\end{equation}
where
\begin{widetext}
\begin{equation}
	\omega(\mathbf{x}) = \mathcal{Z}_{\text{bath}}\frac{(d\tau)^{2k}}{(k!)^2}\left\langle\mathcal{T}_\tau c^{}_{\alpha'_1\sigma'_1}(\tau'_1)c^{\dagger}_{\alpha_1\sigma_1}(\tau_1) \cdots c^{}_{\alpha'_k\sigma'_k}(\tau'_k)c^{\dagger}_{\alpha_k\sigma_k}(\tau_k)\right\rangle_{\mathcal{S}_{\mathrm{loc}}}\det\left[ \underline{\Delta}^{(k)} \right]~.
	\label{}
\end{equation}
\end{widetext}
Here, $\mathbf{x}$ represents the configuration composed of $\left\{k; (\alpha_1,\sigma_1,\tau_1),(\alpha'_1,\sigma'_1,\tau'_1),\dots (\alpha_k,\sigma_k,\tau_k),(\alpha'_k,\sigma'_k,\tau'_k)\right\}$.
One can measure a general time-local observable $\mathcal{O}$ using MC sampling as
\begin{equation}
	\langle \mathcal{O}\rangle_{\text{MC}} = \frac{\sum^{}_{\mathbf{x}}\mathcal{O}(\mathbf{x})\omega(\mathbf{x})}{\sum^{}_{\mathbf{x}}\omega(\mathbf{x})}~.
	\label{}
\end{equation}
A potential sign problem appears when the weight $\omega(\mathbf{x})$ becomes negative for certain configurations.
In this case the observable can be sampled through the modified expression 
\begin{equation}
	\langle \mathcal{O}\rangle_{\text{MC}} = \frac{\sum^{}_{\mathbf{x}}\text{sign}[\omega(\mathbf{x})]~\mathcal{O}(\mathbf{x})~|\omega(\mathbf{x})|}{\sum^{}_{\mathbf{x}}\text{sign}[\omega(\mathbf{x})]~|\omega(\mathbf{x})|}~.
	\label{}
\end{equation}
A severe sign problem occurs if the small average sign
\begin{equation}
	\langle \mathrm{sign}\rangle=\frac{\sum^{}_{\mathbf{x}}\mathrm{sign}[\omega(\mathbf{x})]|\omega(\mathbf{x})|}{\sum^{}_{\mathbf{x}}|\omega(\mathbf{x})|}
	\label{}
\end{equation}
amplifies the error propagation in the estimation of the observables.
Especially, for low temperatures, the relative error of the average sign becomes exponentially large as a function of the inverse temperature $\beta$.
Thus, reliable estimates of observables cannot be obtained in polynomial time.
An analogous phase problem appears if the configuration weights are complex. 
However, since the average phase is always real, we will for simplicity use the terms sign problem and average sign in the following.

An important aspect to keep in mind is that the sign problem depends on the single-particle basis.
For example, it has been reported that if the electron operators $\tilde{c}^{}_{j,m}$ are expressed in the relativistic $j_{\text{eff}}$ basis $V_{j\mathrm{eff}}$,
\begin{eqnarray}
  \left(
  \begin{array}{c}
    \tilde{c}^{}_{\frac{1}{2},+\frac{1}{2}}\\
    \tilde{c}^{}_{\frac{3}{2},+\frac{1}{2}}\\
    \tilde{c}^{}_{\frac{3}{2},-\frac{3}{2}}
  \end{array}
  \right)
  &=& \frac{1}{\sqrt{6}}
  \left(
  \begin{array}{ccc}
	  -\sqrt{2} & +i\sqrt{2} & -\sqrt{2}\\
	  -1 & i & 2 \\
	  -\sqrt{3} & i\sqrt{3} & 0 \\
  \end{array}
  \right)
  \left(
  \begin{array}{c}
	  c^{}_{yz,\downarrow}\\
	  c^{}_{zx,\downarrow}\\
	  c^{}_{xy,\uparrow}\\
  \end{array}
  \right)~,\nonumber\\
  \left(
  \begin{array}{c}
    \tilde{c}^{}_{\frac{1}{2},-\frac{1}{2}}\\
    \tilde{c}^{}_{\frac{3}{2},-\frac{1}{2}}\\
    \tilde{c}^{}_{\frac{3}{2},+\frac{3}{2}}
  \end{array}
  \right)
  &=& \frac{1}{\sqrt{6}}
  \left(
  \begin{array}{ccc}
	  -\sqrt{2} & -i\sqrt{2} & +\sqrt{2}\\
	  +1 & i & 2 \\
	  +\sqrt{3} & i\sqrt{3} & 0 \\
  \end{array}
  \right)
  \left(
  \begin{array}{c}
	  c^{}_{yz,\uparrow}\\
	  c^{}_{zx,\uparrow}\\
	  c^{}_{xy,\downarrow}\\
  \end{array}
  \right),\hspace{8mm}
  \label{eqn:Ujeff}
\end{eqnarray}
the average sign is improved~\cite{Sato:2015hq} due to the diagonalized hybridization function matrix.\cite{Meng:2014jo}  (An average sign of 1 means no sign problem, while an average sign approaching 0 means a severe sign problem.) 

To further optimize the basis beyond $V_{j\mathrm{eff}}$, we introduce the
numerical optimization scheme SPSA. In this scheme the average sign becomes the
objective function on the parameter space where each point corresponds to a
single-particle basis used in CTHYB.
The SPSA approximates the gradient for a given parameter point.
At each iteration, the objective function is measured at both the positively and negatively perturbed parameter points along a stochastically chosen direction.
The gradient is evaluated from this two-point measurement, and the parameter point is updated.
Compared to the finite-difference stochastic approximation which involves a number of measurements proportional to the dimension of the parameter space, the SPSA potentially converges faster in time if the parameter space is high-dimensional and the measurement of the objective function is computationally expensive.

\begin{figure}[t]
	\centering
	\includegraphics[width=0.45\textwidth]{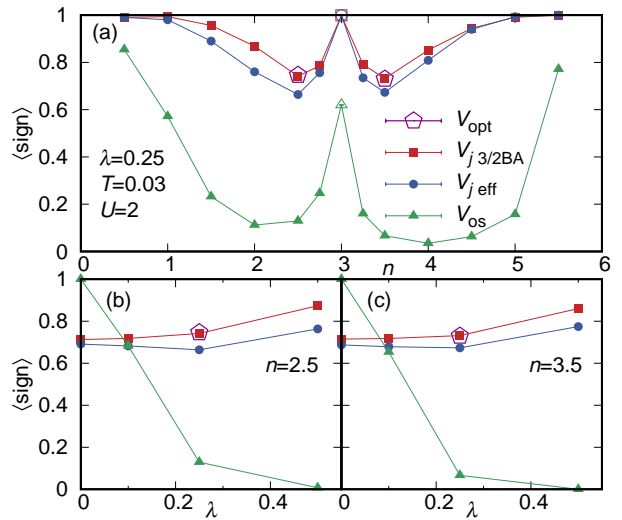}
	\caption{
		Average sign for four different bases as a function of (a) the electron density $n$ and (b) and (c) the strength of SOC $\lambda$ for $T=0.03$, $U=2.0$, and $J_{\mathrm{H}}/U=0.15$~.
		$V_{\mathrm{opt}}$, $V_{j3/2\mathrm{BA}}$, $V_{j \mathrm{eff}}$, and $V_{\mathrm{os}}$ represent, respectively, the optimal basis found by the SPSA, the $j=3/2$ bonding-antibonding basis, the $j_{\mathrm{eff}}$ basis, and the orbital-spin basis.
		The average sign is calculated using the self-consistent paramagnetic solution for a given parameter set.
		The open squares and circles at $n=3$ correspond to the average signs of the Mott insulator solution.
}
	\label{fig:nldep}
\end{figure}
\section{Results}
\begin{figure}[t]
	\centering
	\includegraphics[width=0.45\textwidth]{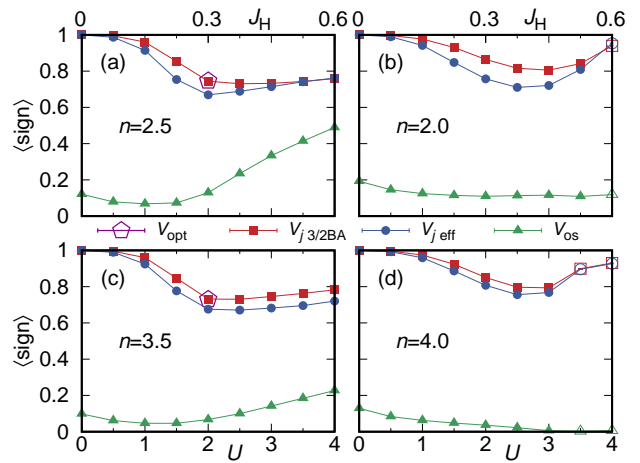}
	\caption{
		Average sign of four different bases as a function of the intraorbital interaction strength $U$ 
		for $T=0.03$, $\lambda=0.25$, and $J_{\mathrm{H}}/U=0.15$. 
		(a)-(d) correspond to the system with electron filling $n=2.5$, $2.0$, $3.5$, and $4.0$, respectively.
		Open squares and circles at $n=2$ and $4$ indicate that the self-consistent paramagnetic solution is Mott insulating.
}
	\label{fig:Udep}
\end{figure}

In Fig.~\ref{fig:nldep} we plot the average sign for the four different bases considered in this work as a function of electron density
[Fig.~\ref{fig:nldep}(a)]  and spin-orbit coupling strength [Fig.~\ref{fig:nldep}(b) and (c)]. 
The orbital-spin basis $V_{\mathrm{os}}$ has a severe sign problem 
since a drastic drop in the average sign of $V_{\mathrm{os}}$ appears as a function of the SOC strength both below and above half-filling. 
Figure~\ref{fig:Udep} shows the evolution of the average sign as a function of $U$ ($J_{\mathrm{H}}$) at various electron fillings.
The suppressed average sign of $V_{\mathrm{os}}$ in the noninteracting limit 
  implies that the source of the sign problem in this basis is the off-diagonal hybridization function generated by the SOC.
The basis $V_{j\mathrm{eff}}$, defined in Eq.~(\ref{eqn:Ujeff}), diagonalizes
the hybridization-function matrix of the Hamiltonian and thus recovers the
sign-free behavior in the noninteracting limit, as demonstrated in Fig.~\ref{fig:Udep}.  In the presence of a nonzero
Hund coupling $J_\mathrm{H}$, however, $V_{j\mathrm{eff}}$ shows a sign problem
as well, especially at intermediate interaction strengths and away from half-filling.

In order to further improve the average sign we introduce the basis $V_{j3/2\mathrm{BA}}$ in which $\tilde{c}^{}_{\frac{3}{2},+\frac{3}{2}}$ and $\tilde{c}^{}_{\frac{3}{2},-\frac{3}{2}}$ are mixed in the bonding-antibonding manner:
\begin{equation}
  \left(
  \begin{array}{c}
	  \tilde{a}^{}_{\frac{3}{2},+}\\
	  \tilde{a}^{}_{\frac{3}{2},-}
  \end{array}
  \right)
  = \frac{1}{\sqrt{2}}
  \left(
  \begin{array}{cccccc}
    1 & 1 \\
    1 & -1 \\
  \end{array}
  \right)
  \left(
  \begin{array}{c}
	  \tilde{c}^{}_{\frac{3}{2},+\frac{3}{2}}\\
	  \tilde{c}^{}_{\frac{3}{2},-\frac{3}{2}}
  \end{array}
  \right)~.\\
  \label{eqn:Uj3/2BA}
\end{equation}
A similar transformation was previously explored for an isolated trimer.\cite{2015PhRvB..92s5126S} It turns out that $V_{j3/2\mathrm{BA}}$ is superior to $V_{j\mathrm{eff}}$ for most parameter sets investigated.
In the noninteracting limit, $V_{j3/2\mathrm{BA}}$ (like $V_{j\mathrm{eff}}$) yields an average sign of unity (see Fig.~\ref{fig:Udep}). 
As $U$ is switched on, $V_{j3/2\mathrm{BA}}$ results in a larger average sign than $V_{j\mathrm{eff}}$~.
In Fig.~\ref{fig:Udep}, the improvement of the average sign in $V_{j3/2\mathrm{BA}}$ is
 prominent in the itinerant regime with intermediate Coulomb interaction strength.
This improvement persists over the whole range of electron densities for $U=2$ as shown in Fig~\ref{fig:nldep}(a)~.
In the Mott insulating regime (marked by open squares and circles in Figs.~\ref{fig:nldep} and \ref{fig:Udep}), 
on the other hand, the difference between the average signs of
$V_{j3/2\mathrm{BA}}$ and $V_{j\mathrm{eff}}$ is smaller.  Since the CTHYB
is based on the expansion around the localized limit, those observations imply
that $V_{j3/2\mathrm{BA}}$ effectively prevents sign-problematic high-order
processes.

Note that neither $V_{j3/2\mathrm{BA}}$ nor $V_{j\mathrm{eff}}$ is an optimal basis for the non-SO-coupled model.
As we show in Figs.~\ref{fig:nldep}(b) and (c), at the
non-SO-coupled point ($\lambda=0$), $V_{j3/2\mathrm{BA}}$ and $V_{j\mathrm{eff}}$
exhibit a sign problem, in contrast to $V_{\mathrm{os}}$~.
This demonstrates that Eq.~(\ref{eqn:kanamori}) includes dangerous interacting terms in the $V_{j3/2\mathrm{BA}}$ and $V_{j\mathrm{eff}}$ basis.
One of these terms is the correlated-hopping (CH) term, which has the form 
\begin{eqnarray}
	&&\frac{\sqrt{2}J_{\text{H}}}{3}\sum^{}_{s=\pm}\left[s(2\tilde{n}_{\frac{3}{2},s\frac{3}{2}} - \tilde{n}_{\frac{3}{2},\bar{s}\frac{3}{2}}- \tilde{n}_{\frac{3}{2},\bar{s}\frac{1}{2}})\tilde{c}^{\dagger}_{\frac{1}{2},s\frac{1}{2}}\tilde{c}^{}_{\frac{3}{2},s\frac{1}{2}} + \text{H.c.}\right]\nonumber\\
	\label{eqn:ch_Ujeff}
\end{eqnarray}
in $V_{j\mathrm{eff}}$ and 
\begin{eqnarray}
	&&\frac{\sqrt{2}J_{\text{H}}}{6}\sum^{}_{s=\pm}\left[ s(\tilde{n}_{\frac{3}{2},+} + \tilde{n}_{\frac{3}{2},-}   - 2 \tilde{n}_{\frac{3}{2},\bar{s}\frac{1}{2}}) \tilde{c}^{\dagger}_{\frac{1}{2},s\frac{1}{2}}\tilde{c}^{}_{\frac{3}{2},s\frac{1}{2}}+\text{H.c.}\right]\nonumber\\
	&&+\frac{J_{\mathrm{H}}}{2}\left[(\tilde{n}_{\frac{1}{2},+\frac{1}{2}}-\tilde{n}_{\frac{1}{2},-\frac{1}{2}})\tilde{a}^{\dagger}_{\frac{3}{2},+}a^{}_{\frac{3}{2},-} + \text{H.c.}\right]
	\label{eqn:ch_Uj32BA}
\end{eqnarray}
in $V_{j3/2\mathrm{BA}}$~.
This CH term is the major source of the sign problem in the SO-coupled Hamiltonian.
Table~\ref{tab:mask} analyzes the effect of different terms in the local
 Hamiltonian on the average sign for both $V_{j3/2\mathrm{BA}}$ and $V_{j\mathrm{eff}}$~.
There is a substantial drop in the average sign when the CH terms are introduced.

\begin{table}
	\centering
	\begin{tabular}{p{0.55\linewidth}p{0.20\linewidth}p{0.20\linewidth}}
		\hline\hline
		Terms& $\langle\mathrm{sign}\rangle_{V_{j\mathrm{eff}}}$ & $\langle\mathrm{sign}\rangle_{V_{j3/2\mathrm{BA}}}$\vspace{0.5pt}\\
		\hline
		Density-density (DD) & 1.00000 & 1.00000\\
		DD + spin flip (SF) & 1.00000 & 1.00000\\
		DD + four scattering (FS) & 0.99447(9) & 0.9516(3)\\
		DD + SF + pair hopping (PH) & 0.9760(2) & 0.9739(2)\\
		DD + PH & 0.9730(7) & 0.9705(2)\\
		DD + SF + PH + FS & 0.9308(3) & 0.9035(6)\\
		DD + \textit{correlated hopping} (CH) & 0.8212(9) & 0.8730(5)\\
		DD + CH + FS & 0.8061(6) & 0.7718(5)\\
		DD + PH + CH & 0.7728(6) & 0.8623(6)\\
		DD + SF + PH + CH & 0.7638(7) & 0.8814(5)\\
		DD + SF + PH + CH + FS & 0.6751(8) & 0.7406(4)\\
		\hline\hline
	\end{tabular}
	\caption{
		Average sign for the masked local Hamiltonians represented in $V_{j\mathrm{eff}}$ and $V_{j3/2\mathrm{BA}}$~.
		The results are for $n=3.5$, $T=0.03$, $\lambda=0.25$, $U=2.0$, and $J_\mathrm{H}/U$=0.15, and the 
		self-consistent hybridization for the full local Hamiltonian. 
	}
	\label{tab:mask}
\end{table}

In $V_{j3/2\mathrm{BA}}$ and $V_{j\mathrm{eff}}$, there emerge other new terms involving four different flavors, which are not of the spin-flip type or pair-hopping type appearing in the Slater-Kanamori Hamiltonian in $V_{\mathrm{os}}$~.
Those terms are denoted as four-scattering terms in Table~\ref{tab:mask}~.
The average sign in both $V_{j3/2\mathrm{BA}}$ and $V_{j\mathrm{eff}}$ becomes even lower when the CH term is combined with the pair-hopping and the four-scattering terms.
The full interacting Hamiltonian in $V_{j\text{eff}}$ is explicitly written in Appendix~\ref{app:HintVjeff}.
Furthermore, we analyze the nature of the self-consistent solutions for the Hamiltonians without the CH or FS terms and we discuss the potential use of such {\it masked} Hamiltonians in Appendix~\ref{app:maskH}.

\begin{figure}[]
	\centering
	\includegraphics[width=0.45\textwidth]{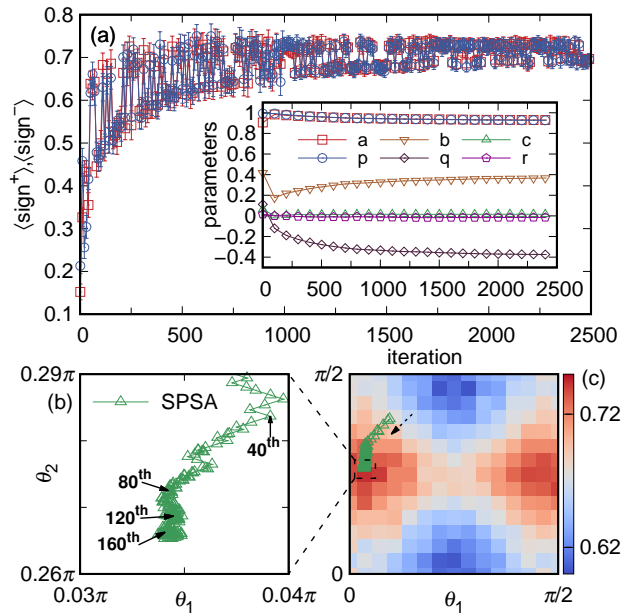}
	\caption{
		(a) Average sign of the positively and negatively perturbed parameter points during the SPSA search for the $SO(4)$ group ($n=3.5$, $T=0.03$, $\lambda=0.25$, $U=2.0$, and $J_{\mathrm{H}}/U=0.15$) as a function of iterations.
		The inset shows the convergence of the parameters defined in Eqs.~(\ref{eqn:ML}).
		(c) Landscape of the average sign in the $(\theta_1,\theta_2)$ subspace described by Eq.~(\ref{eqn:theta12space}).
		Green triangles represent the SPSA trajectory in the parameter space.
		(b) Enlarged plot of the SPSA trajectory of (c).
		The numeric labels in (b) denote iteration numbers during the SPSA search, and the arrow in (c) shows the direction in which the search proceeds.
	}
	\label{fig:SPSA}
\end{figure}

In what follows we will determine by the SPSA method the optimal basis in terms of the average sign for the
parameters with the most severe sign problem and show
that this optimal basis is nearly identical to the $V_{j3/2\mathrm{BA}}$ basis.
For that,
we search the basis space generated by the $SO(4)$ rotation group for $j=3/2$, whose $4\times 4$ matrix representation is denoted by $\mathbbm{M}$ and transforms the electron operators in $V_{j\mathrm{eff}}$ as follows: 
\begin{gather}
	\tilde{a}^{}_{\frac{3}{2}} = \mathbbm{M}\cdot\tilde{c}^{}_{\frac{3}{2}}~,\\
	\tilde{a}^{\intercal}_{\frac{3}{2}} = \left(\tilde{a}^{}_{\frac{3}{2},1},\tilde{a}^{}_{\frac{3}{2},2},\tilde{a}^{}_{\frac{3}{2},3},\tilde{a}^{}_{\frac{3}{2},4}\right)~,\\
	\tilde{c}^{\intercal}_{\frac{3}{2}} = \left(\tilde{c}^{}_{\frac{3}{2},+\frac{1}{2}},\tilde{c}^{}_{\frac{3}{2},-\frac{1}{2}},\tilde{c}^{}_{\frac{3}{2},+\frac{3}{2}},\tilde{c}^{}_{\frac{3}{2},-\frac{3}{2}}\right)~.
	\label{}
\end{gather}
Since the off-diagonal hybridization function is a clear source of the severe sign problem as shown in the cases of $V_{\mathrm{os}}$ (Figs.~\ref{fig:nldep} and \ref{fig:Udep}), we exclude the mixing between the $j=1/2$ and $3/2$ subspaces to preserve the diagonal structure of the hybridization.
Without mixing between the $j=1/2$ and $3/2$ subspaces, one can fix the basis for the $j=1/2$ subspace using the rotational symmetry generated by the total angular momentum operator, $\mathbf{J}_{\mathrm{eff}}=\mathbf{L}_{\mathrm{eff}}+\mathbf{S}$ without loss of generality.

To parametrize the basis space, we introduce the isoclinic decomposition of $\mathbbm{M}$ as
	$\mathbbm{M} = \mathbbm{M}_{L}\mathbbm{M}_{R}$~,
where
\begin{align}
	&\mathbbm{M}_{L} {=}
	\left(
	\begin{array}{cccc}
		a & -b & -c & -d\\
		b & a & -d & c\\
		c & d & a & -b\\
		d & -c & b & a
	\end{array}
	\right),\hspace{2mm}
	\mathbbm{M}_{R} 
	{=}
	\left(
	\begin{array}{cccc}
		p & -q & -r & -s\\
		q & p & s & -r\\
		r & -s & p & q\\
		s & r & -q & p
	\end{array}
	\right).
		\label{eqn:ML}
\end{align}
Here, $a^2+b^2+c^2+d^2 = 1$ and $p^2+q^2+r^2+s^2 = 1$~.
Under these constraints among $\{a,b,c,d\}$ and $\{p,q,r,s\}$ the dimension of the parameter space becomes $6$.

Figure~\ref{fig:SPSA} shows how the SPSA works while searching for the optimal basis in this parameter space.
The evolution of the average sign as a function of number of iterations
at both positively ($\langle\mathrm{sign}^+\rangle$) and negatively ($\langle\mathrm{sign}^-\rangle$) perturbed points in the parameter space is shown in Fig.~\ref{fig:SPSA} (a).
The average sign value converges to $\sim 0.74$ for $n=3.5$, $T=0.03$, $\lambda=0.25$, $U=2.0$, and $J_\mathrm{H}/U=0.15$~.
Within numerical accuracy, it is very close to the value of
$V_{j3/2\mathrm{BA}}$ defined in Eq.~(\ref{eqn:Uj3/2BA})~.  This shows that the
$V_{j3/2\mathrm{BA}}$ basis is at least near the local optimum in parameter
space. 
The inset of Fig.~\ref{fig:SPSA}(a) illustrates the convergence of the parameters in Eqs.~(\ref{eqn:ML}). 

Figure~\ref{fig:SPSA}(b) and (c) show the SPSA sequence in a small parameter subspace. 
We introduce two parameters, $\theta_1$ and $\theta_2$, representing the restricted basis transformation 
\begin{equation}
	\mathbbm{M}
	= \left(
	\begin{array}{cccc}
		\cos\theta_1 & \sin\theta_1 & 0 & 0\\
		-\sin\theta_1 & \cos\theta_1 & 0 & 0\\
		0 & 0 & \cos\theta_2 & \sin\theta_2\\
		0 & 0 & -\sin\theta_2 & \cos\theta_2\\
	\end{array}
	\right)~.
	\label{eqn:theta12space}
\end{equation}
 Figure~\ref{fig:SPSA}(c) shows that the landscape of the average sign is smooth, so that the
SPSA search based on the gradient approximation can successfully find the local optimum in that subspace.
Furthermore, $V_{j3/2\mathrm{BA}}$, corresponding to $\theta_1=0$ and $\theta_2=\pi/4$, is shown to be very close to the optimum found by the SPSA.
Figure~\ref{fig:SPSA}(b) plots the trajectory determined by the SPSA.

\begin{figure}[]
	\centering
	\includegraphics[width=0.45\textwidth]{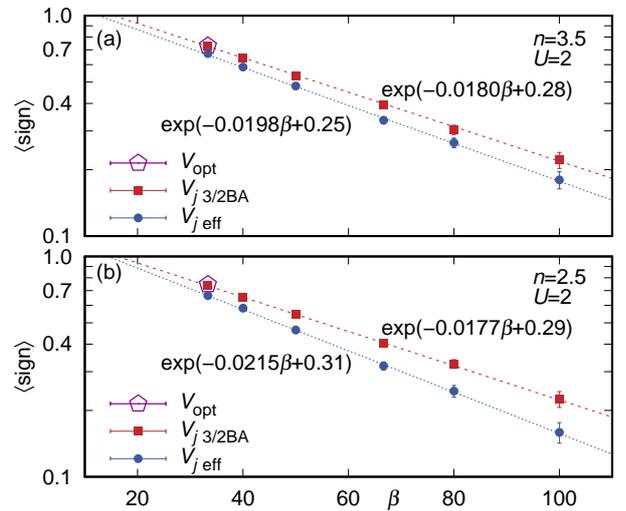}
	\caption{
		Temperature dependence of the average sign 
		for $T=0.03$, $\lambda=0.15$, $U=2.0$, and $J_{\mathrm{H}}/U=0.15$~. 
		The electron filling for (a) and (b) is $3.5$ and $2.5$, respectively.
		The dashed (dotted) line corresponds to the fitting function of the form $\exp\left[-\beta\delta F + F_0\right]$ for the $V_{j3/2\mathrm{BA}}$ ($V_{j{\mathrm{eff}}}$) basis.
	}
	\label{fig:Tdep}
\end{figure}

We finally investigate the temperature scaling of the average sign for the different bases.
Figure~\ref{fig:Tdep} shows the exponential scaling $\langle \mathrm{sign}\rangle \sim \exp\left(-\beta\delta F + F_0\right)$ 
as a function of the inverse temperature $\beta$. 
Here, $\delta F$ can be regarded as the free-energy difference from the auxiliary bosonic system without a sign problem, which determines the temperature scaling. 
For both fillings, $n=3.5$ and $2.5$, the $V_{j3/2\mathrm{BA}}$ basis shows an improved temperature scaling exponent $\delta F$ and a comparable offset $F_0$~.

\section{Conclusion}
We have investigated the nature of the sign problem in the CTHYB for the SO-coupled multiorbital Hubbard model.
We found that the correlated hopping term that
appears in the $j_{\mathrm{eff}}$ basis is the major source of the sign problem,
 and we introduced a basis--the $j=3/2$ bonding-antibonding basis--which alleviates the effects of those terms.
By applying the stochastic optimization scheme, we found that
the $j=3/2$ bonding-antibonding basis is near a local optimum in the parameter space considered.
These results (i) provide useful guidelines for the choice
 of the single-particle basis in CTHYB simulations of SO-coupled systems,
(ii) introduce an algorithm to numerically determine the optimal basis, and 
(iii) suggest using the intrinsic fermionic sign bound in a
systematic search of the full $SO(4)$ basis space
as a strategy for the further improvement of CTHYB for this class of systems. 

In addition, our findings have relevance for descriptions beyond DMFT that include
nonlocal correlations. For example, 
since the average sign is alleviated at the level of the partition function in $V_{j3/2\text{BA}}$,
we expect our results to be helpful for a reliable estimation of two-particle correlation functions,
which are the essential building blocks of diagrammatic extensions of DMFT.\cite{Rohringer2018} 
For cluster extensions of DMFT, on the other hand, our numerical basis optimization scheme can be applied to an extended basis space including site (cluster) degrees of freedom. 
A promising strategy might be to combine the $V_{j3/2\text{BA}}$ basis with a bonding-antibonding transformation that breaks loops on the cluster.\cite{2015PhRvB..92s5126S}

{\it Note added.} During the final stages of this work, a series of conceptually related studies appeared on  arXiv.\cite{Hangleiter2019,Torlai2019,Levy2019}

\begin{acknowledgments}
A.J.K. was supported by EPSRC through Grant No. EP/P003052/1. 
P.W. acknowledges support from the Swiss National Science Foundation via NCCR Marvel. 
R.V. acknowledges support by the Deutsche Forschungsgemeinschaft (DFG) 
through Grant No. VA117/15-1. 
The computations were performed at the Centre for Scientific Computing (CSC)
in Frankfurt.
\end{acknowledgments}

\bibliography{ref}

\appendix
\section{Energy scales in materials}
\label{app:material}
In this appendix, we list materials related to our spin-orbit-coupled (SOC) three-orbital model, and the corresponding energy scales.  
The SOC is an intrinsic relativistic effect of the atom whose strength is an increasing function of the atomic number.
Hence, $5d$ materials (Ta, W, Os, and Ir) consistently show a larger SOC strength compared to $4d$ materials (Ru and Rh).
Table~\ref{tab:material} lists materials with various electron configurations which correspond to the electron density in our model.
The ratio between the SOC strength and the half-bandwidth used in our model is given in the fourth column.
\begin{table}[b]
  \centering
  \begin{tabular}{ccccc}
    \hline\hline
    Material & Conf. & $D$ (eV) & $\lambda/D$ & Ref.\\
    \hline
    Sr$_2$IrO$_4$ & \multirow{2}{*}{$d^5$} & \multirow{2}{*}{1.44} & 0.257 (0.37 eV) & \multirow{2}{*}{\cite{Watanabe:2010}}\\
    Sr$_2$RhO$_4$ &  & & 0.125 (0.18 eV) & \\
    \hline
    Y$_2$Ir$_2$O$_7$ & $d^5$ & 0.5 & 0.8 (0.4 eV) & \cite{Shinaoka:2015cg}\\
    \hline
    Na$_2$IrO$_3$ & \multirow{3}{*}{$d^5$} & \multirow{3}{*}{0.2} & \multirow{2}{*}{2.0 (0.4 eV)} & \multirow{3}{*}{\cite{Winter:2016}}\\
    Li$_2$IrO$_3$ &  &  &  & \\
    $\alpha$-RuCl$_3$ &  &  & 0.75 (0.15 eV) & \\
    \hline
    GaW$_4$Se$_4$Te$_4$ & $d^5$ & 0.195 & 1.23 (0.24 eV) & \cite{Kim2014}\\
    \hline
    Sr$_2$RuO$_4$ & $d^4$ & 0.75 & 0.13 (0.100 eV) & \cite{Haverkort:2008}\\
    \hline
    Sr$_2$YRuO$_6$ & $d^4$ & 0.55 & 0.18 (0.100 eV) & \cite{Meetei:2015,Vaugier:2012,Mazin:1997}\\
    \hline
    Sr$_2$YIrO$_6$ & \multirow{2}{*}{$d^4$} & \multirow{2}{*}{0.5} & \multirow{2}{*}{0.6 (0.33 eV)} & \multirow{2}{*}{\cite{Pajskr:2016}}\\
    Ba$_2$YIrO$_6$ &  & & & \\
    \hline
    Sr$_2$CrOsO$_6$ & $d^3$ & 0.75 & 0.4 (0.3 eV) & \cite{Meetei:2013}\\
    \hline
    Ba$_2$NaOsO$_6$  & $d^1$ & 0.3 & 1.0 (0.3 eV) & \cite{Gangopadhyay:2016,Lee:2007}\\
    \hline
    GaTa$_4$Se$_8$  & $d^1$ & 0.38 & 0.55 (0.21 eV) & \cite{Kim2014}\\
    \hline
    GaTa$_4$Se$_4$Te$_4$ & $d^1$ & 0.22 & 0.95 (0.21 eV) & \cite{Kim2014}\\
    \hline\hline
  \end{tabular}
  \caption{
	  Summary of the approximate energy scales in various SOC materials. Conf., $D$, and $\lambda$ present the electronic configuration, the half-bandwidth, and strength of spin-orbit coupling, respectively.  The actual SOC strength is written in parentheses in the fourth column.
  }
  \label{tab:material}
\end{table}

Since the proposed $V_{j3/2\mathrm{BA}}$ basis alleviates the sign problem in a wide parameter range (electron density and strength of the SOC and many-body interaction), as shown in Figs. 1 and 2 of the main text, it should enable improved material-specific calculations for a broad range of materials.

\section{Interacting Hamiltonian in the $V_{j\text{eff}}$ basis}
\label{app:HintVjeff}
In Eq.~(\ref{eqn:HintVjeff}), we classify the interacting Hamiltonian in the $V_{j\mathrm{eff}}$ basis into five classes: density-density ($\mathcal{H}_{\text{dd}}$), correlated-hopping ($\mathcal{H}_{\text{ch}}$), pair-hopping ($\mathcal{H}_{\text{ph}}$), four-scattering ($\mathcal{H}_{\text{fs}}$), and spin-flip ($\mathcal{H}_{\text{sf}}$) terms.
The Hamiltonian in the $V_{j3/2\mathrm{BA}}$ basis is classified in the same way.
\begin{widetext}
\begin{eqnarray}
	\mathcal{H}_{\text{int}} &=& \mathcal{H}_{\text{dd}} + \mathcal{H}_{\text{ch}} + \mathcal{H}_{\text{ph}} + \mathcal{H}_{\text{fs}} + \mathcal{H}_{\text{sf}}~,\nonumber\\
	\mathcal{H}_{\text{dd}} &=& (U- \frac{4J_{\text{H}}}{3})\tilde{n}_{\frac{k}{2},+\frac{1}{2}}\tilde{n}_{\frac{1}{2},-\frac{1}{2}} + (U-J_{\text{H}})(\tilde{n}_{\frac{3}{2},\frac{1}{2}}\tilde{n}_{\frac{3}{2},-\frac{1}{2}} + \tilde{n}_{\frac{3}{2},\frac{3}{2}}\tilde{n}_{\frac{3}{2},-\frac{3}{2}}) \nonumber\\
	&&+ (U-\frac{7J_{\text{H}}}{3})\sum^{}_{s=\pm}(\tilde{n}_{\frac{1}{2},s\frac{1}{2}}\tilde{n}_{\frac{3}{2},\bar{s}\frac{1}{2}} + \tilde{n}_{\frac{3}{2},s\frac{1}{2}}\tilde{n}_{\frac{3}{2},s\frac{3}{2}} + \tilde{n}_{\frac{3}{2},s\frac{1}{2}}\tilde{n}_{\frac{3}{2},\bar{s}\frac{3}{2}})\nonumber\\
	&&+ (U-2J_{\text{H}})\sum^{}_{s=\pm}\tilde{n}_{\frac{1}{2},s\frac{1}{2}}\tilde{n}_{\frac{3}{2},s\frac{1}{2}} + (U-\frac{8J_{\text{H}}}{3})\sum^{}_{s=\pm}\tilde{n}_{\frac{1}{2},s\frac{1}{2}}\tilde{n}_{\frac{3}{2},\bar{s}\frac{3}{2}} + (U-\frac{5J_{\text{H}}}{3})\sum^{}_{s=\pm}\tilde{n}_{\frac{1}{2},s\frac{1}{2}}\tilde{n}_{\frac{3}{2},s\frac{3}{2}}~,\nonumber\\
	\mathcal{H}_{\text{ch}} & = &\frac{\sqrt{2}J_{\text{H}}}{3}\sum^{}_{s=\pm}\left[s(2\tilde{n}_{\frac{3}{2},s\frac{3}{2}} - \tilde{n}_{\frac{3}{2},\bar{s}\frac{3}{2}}- \tilde{n}_{\frac{3}{2},\bar{s}\frac{1}{2}})\tilde{c}^{\dagger}_{\frac{1}{2},s\frac{1}{2}}\tilde{c}^{}_{\frac{3}{2},s\frac{1}{2}} + \text{H.c.}\right]~,\nonumber\\
	\mathcal{H}_{\text{ph}} &=&- \frac{5J_{\text{H}}}{3}\left[\tilde{c}^{\dagger}_{\frac{1}{2},+\frac{1}{2}}\tilde{c}^{\dagger}_{\frac{1}{2},-\frac{1}{2}}(\tilde{c}^{}_{\frac{3}{2},-\frac{1}{2}}\tilde{c}^{}_{\frac{3}{2},+\frac{1}{2}} - \tilde{c}^{}_{\frac{3}{2},-\frac{3}{2}}\tilde{c}^{}_{\frac{3}{2},+\frac{3}{2}}) + \text{H.c.}\right] - \frac{4J_{\text{H}}}{3}\left[\tilde{c}^{\dagger}_{\frac{3}{2},+\frac{1}{2}}\tilde{c}^{\dagger}_{\frac{3}{2},-\frac{1}{2}}\tilde{c}^{}_{\frac{3}{2},-\frac{3}{2}}\tilde{c}^{}_{\frac{3}{2},+\frac{3}{2}} + \text{H.c.}\right]~,\nonumber\\
	\mathcal{H}_{\text{fs}}&=&- \frac{J_{\text{H}}}{\sqrt{3}}\sum^{}_{s=\pm}\left[\tilde{c}^{\dagger}_{\frac{1}{2},s\frac{1}{2}}\tilde{c}^{\dagger}_{\frac{3}{2},s\frac{1}{2}}\tilde{c}^{}_{\frac{1}{2},\bar{s}\frac{1}{2}}\tilde{c}^{}_{\frac{3}{2},s\frac{3}{2}}  + \tilde{c}^{\dagger}_{\frac{1}{2},s\frac{1}{2}}\tilde{c}^{\dagger}_{\frac{3}{2},\bar{s}\frac{3}{2}}\tilde{c}^{}_{\frac{1}{2},\bar{s}\frac{1}{2}}\tilde{c}^{}_{\frac{3}{2},\bar{s}\frac{1}{2}} + \text{H.c.}\right]\nonumber\\
	&&- \frac{\sqrt{2}J_{\text{H}}}{\sqrt{3}}\sum^{}_{s=\pm}s\left[\tilde{c}^{\dagger}_{\frac{1}{2},s\frac{1}{2}}\tilde{c}^{\dagger}_{\frac{3}{2},\bar{s}\frac{1}{2}}\tilde{c}^{}_{\frac{3}{2},\bar{s}\frac{3}{2}}\tilde{c}^{}_{\frac{3}{2},s\frac{3}{2}} + \tilde{c}^{\dagger}_{\frac{1}{2},s\frac{1}{2}}\tilde{c}^{\dagger}_{\frac{3}{2},s\frac{1}{2}}\tilde{c}^{}_{\frac{3}{2},\bar{s}\frac{1}{2}}\tilde{c}^{}_{\frac{3}{2},s\frac{3}{2}} + \text{H.c.}\right]~,\nonumber\\
	\mathcal{H}_{\text{sf}}&=& \frac{2J_{\text{H}}}{3}\sum^{}_{s=\pm}\left[\tilde{c}^{\dagger}_{\frac{1}{2},s\frac{1}{2}}\tilde{c}^{\dagger}_{\frac{3}{2},\bar{s}\frac{1}{2}}\tilde{c}^{}_{\frac{3}{2},s\frac{1}{2}}\tilde{c}^{}_{\frac{1}{2},\bar{s}\frac{1}{2}} + \text{H.c.}\right]~.
\label{eqn:HintVjeff}
\end{eqnarray}
\end{widetext}

\section{Self-consistent solution of the masked Hamiltonians}
\label{app:maskH}
In this appendix, we investigate the nature of the self-consistent solutions for masked Hamiltonians.
Since we can increase the average sign substantially by dropping the most problematic correlated-hopping (CH) or four-scattering (FS) terms, 
those masked Hamiltonians potentially provide a useful approximation of the full Hamiltonian if the self-consistent solution is sufficiently close to the one for the full Hamiltonian.

The masking of those terms, however, modifies the local eigenstates.
Table~\ref{tab:GS} shows the form of the ground states of the full local Hamiltonian in $V_{j\mathrm{eff}}$ defined in the main text.
When we mask the CH or FS terms, the ground-state degeneracy of the full local Hamiltonian is broken in the $N=2$ and $3$ sectors.
Tables~\ref{tab:GS_wo_CH} and \ref{tab:GS_wo_FS} present the ground states for the masked local Hamiltonians without the CH and FS terms, respectively.
The remaining ground states for the $N=2$ and $3$ sectors depend on the type of the masked terms.
When the FS terms are dropped, the form of the highest $|J_z|$ ground states remains the same as that for the full Hamiltonian.
Since the FS terms involve four different flavors by definition, they become irrelevant for the highest $|J_z|$ states with fixed $j=3/2$ and $m_j=\pm 3/2$ electrons.
On the other hand, masking the CH terms selects the $|J_z|=1$ states for the $N=2$ sector and the $|J_z|=1/2$ states for the $N=3$ sector with slightly modified coefficients, which demonstrates the relevance of these terms for the highest $|J_z|$ states.

\begin{figure*}[]
	\centering
	\includegraphics[width=1.0\textwidth]{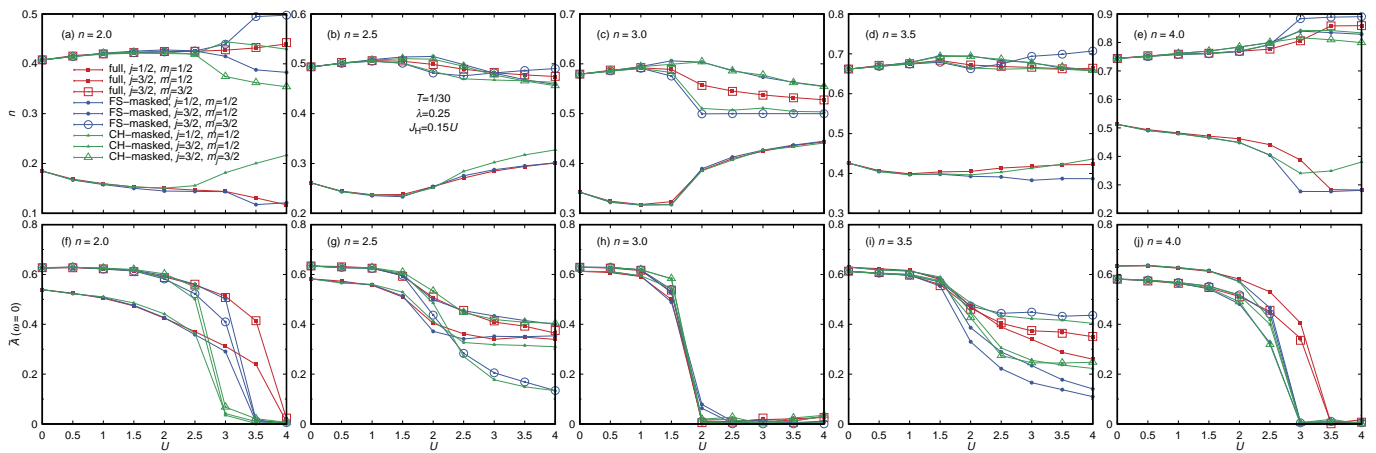}
	\caption{
		(a)-(e) Electron density and (f)-(j) approximate spectral function at the Fermi level as a function of interaction strength $U$ for various total electron fillings.
		Both observables are measured from the self-consistent solutions of the full and masked Hamiltonians.
		For the masked Hamiltonians, the FS and CH terms are dropped in the $V_{j\mathrm{eff}}$ basis.
	}

	\label{fig:observables}
\end{figure*}

\begin{figure*}[]
	\centering
	\includegraphics[width=0.4\textwidth]{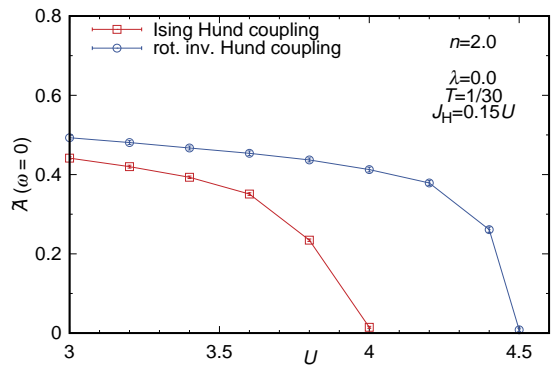}
	\caption{
		Approximate spectral function at the Fermi level for the non-spin-orbit-coupled three-orbital system with Ising-type Hund coupling.  
		Compared to the rotationally-invariant Hund coupling, the Mott transition point $U_c$ is suppressed by $\sim 10\%$ for $n=2$, $T=1/30$, and $J_\mathrm{H}=0.15U$~.
	}
	\label{fig:IsingHund}
\end{figure*}

Such a modified degeneracy of the local Hamiltonian leads to substantial changes in observables, especially when the system becomes localized.
In Figs.~\ref{fig:observables}(a), \ref{fig:observables}(c), and \ref{fig:observables}(e), for example, one can see a sizable difference between the electron densities from the full and masked local Hamiltonians for the large-$U$ Mott insulator.
The degeneracy of the electron density between the $j=3/2$, $m_j=\pm 1/2$, and $m_j=\pm 3/2$ flavors is naturally broken for the masked Hamiltonian.
Mott localization is signaled by the suppression of the spectral function at the Fermi level, and this quantity can be approximately evaluated as $\tilde{A}(\omega=0)=-\frac{1}{\pi T} G(\tau=\beta/2)$~.
Moreover, the Mott transition point $U_c$ for $n=2$ and $4$ is reduced as we mask the CH or FS terms.
Compared to the FS-dropped Hamiltonian, the CH-dropped one shows a further reduction in $U_c$ for the $n=2$ case.
This kind of $U_c$ reduction as a result of a degeneracy breaking was reported in a non-spin-orbit-coupled system in the presence of a single-particle crystal-field splitting~\cite{Werner:2009kc,Huang:2012es} and in the absence of the spin-flip and pair-hopping terms (so-called Ising-type Hund coupling) at the many-body level~\cite{Huang:2012es} (see Fig.~\ref{fig:IsingHund}).

\begin{figure*}[]
	\centering
	\includegraphics[width=1.0\textwidth]{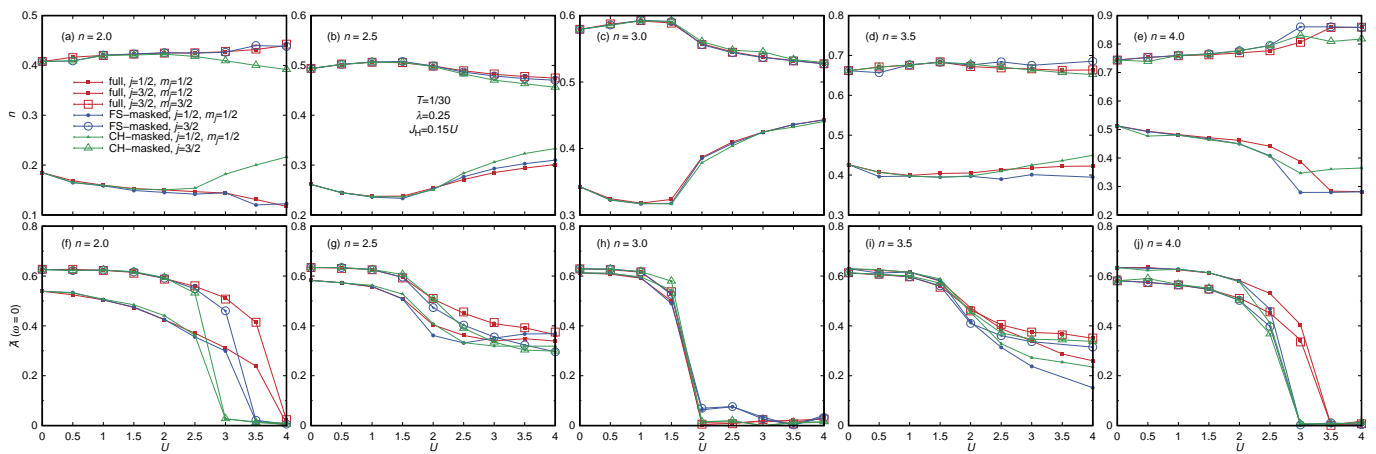}
	\caption{
		(a)-(e) Electron density and (f)-(j) the approximate spectral function at the Fermi level as a function of interaction strength $U$ for various total electron fillings.
		Both observables are measured from the \textit{symmetrized} self-consistent solutions of the full and masked Hamiltonians.
		The symmetrization between the $j=3/2$, $m_j=\pm 1/2$, and $m_j=\pm 3/2$ Green's functions is done at every DMFT iteration step.
		For the masked Hamiltonians, the FS and CH terms are dropped in the $V_{j\mathrm{eff}}$ basis.
	}
	\label{fig:observables_sym}
\end{figure*}

One interesting observation is that the density values of the full Hamiltonian are approximately recovered by the masked one if we artificially symmetrize the $j=3/2$, $m_j=\pm 1/2$, and $m_j=\pm 3/2$ flavors during the self-consistency loop of the DMFT.
Figure~\ref{fig:observables_sym} shows the corresponding electron density and the approximated spectral function at the Fermi level. The density values become much closer to those of the full Hamiltonian.
Especially, the FS-dropped Hamiltonian at and below half-filling yields results which are consistent within $\sim 5\%$ relative error.
As a result of the modification of the Hamiltonian, the sign problem of the CTHYB simulation is alleviated by up to $\sim 15\%$. When applied with proper care, this symmetrization trick could provide a useful estimate for physical observables when the original Hamiltonian cannot be treated due to the serious sign problem.

\begin{table*}[]
 \centering
\begin{tabular}{ccc}
 \hline\hline
 $N$ & $J_z$ & Ground State\\
 \hline
 \multirow{5}{*}{2}
 & +2 & $\alpha_2\LocalState{0}{0}{1}{0}{1}{0} - \beta_2\LocalState{1}{0}{0}{0}{1}{0}$ \\
 & +1 & $\alpha_2'\LocalState{0}{0}{0}{1}{1}{0} - \beta_2'\LocalState{1}{0}{1}{0}{0}{0} - \gamma_2'\LocalState{0}{1}{0}{0}{1}{0}$ \\
 & 0  & $\alpha_2''\left(\LocalState{0}{0}{1}{1}{0}{0} + \LocalState{0}{0}{0}{0}{1}{1}\right) + \beta_2''\left(\LocalState{1}{0}{0}{1}{0}{0} - \LocalState{0}{1}{1}{0}{0}{0}\right) $ \\
 & -1 & $\alpha_2'\LocalState{0}{0}{1}{0}{0}{1} + \beta_2'\LocalState{0}{1}{0}{1}{0}{0} + \gamma_2'\LocalState{1}{0}{0}{0}{0}{1}$ \\
 & -2 & $\alpha_2\LocalState{0}{0}{0}{1}{0}{1} + \beta_2\LocalState{0}{1}{0}{0}{0}{1}$ \\
 \hline
 \multirow{4}{*}{3}
 & +3/2 & $\alpha_3\LocalState{0}{1}{1}{0}{1}{0} - \beta_3\LocalState{0}{0}{1}{1}{1}{0}-\gamma_3\LocalState{1}{1}{0}{0}{1}{0} + \delta_3\LocalState{1}{0}{0}{1}{1}{0}$ \\
 & +1/2 & $\alpha_3'\LocalState{0}{1}{0}{1}{1}{0} - \beta_3'\LocalState{0}{0}{1}{0}{1}{1} - \gamma_3'\LocalState{1}{1}{1}{0}{0}{0} + \delta_3'\left(\LocalState{1}{0}{1}{1}{0}{0} + \LocalState{1}{0}{0}{0}{1}{1}\right)$ \\
 & -1/2 & $\alpha_3'\LocalState{1}{0}{1}{0}{0}{1} - \beta_3'\LocalState{0}{0}{0}{1}{1}{1} + \gamma_3'\LocalState{1}{1}{0}{1}{0}{0} + \delta_3'\left(\LocalState{0}{1}{1}{1}{0}{0} - \LocalState{0}{1}{0}{0}{1}{1}\right)$ \\
 & -3/2 & $\alpha_3\LocalState{1}{0}{0}{1}{0}{1} + \beta_3\LocalState{0}{0}{1}{1}{0}{1} + \gamma_3\LocalState{1}{1}{0}{0}{0}{1} + \delta_3\LocalState{0}{1}{1}{0}{0}{1}$ \\
 \hline
 4 & 0 & $\alpha_4\LocalState{0}{0}{1}{1}{1}{1} + \beta_4\left(\LocalState{1}{1}{0}{0}{1}{1} + \LocalState{1}{1}{1}{1}{0}{0}\right)$\\
 \hline\hline
\end{tabular}
\caption{
	Ground state for a given sector of the \emph{full local Hamiltonian}.
	In our notation, the upper (lower) level represents the $j=1/2$ ($3/2$) flavors, and the lower left (right) level corresponds to the $m_j=\pm 1/2$ ($m_j=\pm 3/2$) flavor.
	Solid (open) circles mark the positive (negative) $m_j$ electron.
	The subscript of the coefficients represents the corresponding $N$ sector.
}
\label{tab:GS}
\end{table*}

\begin{table*}[]
 \centering
\begin{tabular}{ccc}
 \hline\hline
 $N$ & $J_z$ & Ground State\\
 \hline
 \multirow{2}{*}{2}
 & +1 & $\alpha_2'''\LocalState{0}{0}{0}{1}{1}{0} - \beta_2'''\LocalState{1}{0}{1}{0}{0}{0} + \gamma_2'''\LocalState{0}{1}{0}{0}{1}{0}$ \\
 & -1 & $\alpha_2'''\LocalState{0}{0}{1}{0}{0}{1} + \beta_2'''\LocalState{0}{1}{0}{1}{0}{0} - \gamma_2'''\LocalState{1}{0}{0}{0}{0}{1}$ \\
 \hline
 \multirow{2}{*}{3}
 & +1/2 & $\alpha_3''\LocalState{0}{0}{1}{0}{1}{1} - \beta_3''\LocalState{0}{1}{0}{1}{1}{0} + \gamma_3''\LocalState{1}{1}{1}{0}{0}{0} - \delta_3''\LocalState{1}{0}{1}{1}{0}{0} - \eta\LocalState{1}{0}{0}{0}{1}{1}$ \\
 & -1/2 & $\alpha_3''\LocalState{0}{0}{0}{1}{1}{1} - \beta_3''\LocalState{1}{0}{1}{0}{0}{1} - \gamma_3''\LocalState{1}{1}{0}{1}{0}{0} - \delta_3''\LocalState{0}{1}{1}{1}{0}{0} + \eta\LocalState{0}{1}{0}{0}{1}{1}$ \\
 \hline
 4 & 0 & $\alpha_4'\LocalState{0}{0}{1}{1}{1}{1} + \beta_4'\LocalState{1}{1}{1}{1}{0}{0} + \gamma_4'\LocalState{1}{1}{0}{0}{1}{1} + \delta_4'\left(\LocalState{1}{0}{0}{1}{1}{1} + \LocalState{0}{1}{1}{0}{1}{1}\right)$\\
 \hline\hline
\end{tabular}
\caption{
	Ground state for a given sector of the local Hamiltonian \emph{without the correlated-hopping terms}.
	In our notation, the upper (lower) level represents the $j=1/2$ ($3/2$) flavors and the lower left (right) level corresponds to the $m_j=\pm 1/2$ ($m_j=\pm 3/2$) flavor.
	Solid (open) circles mark the positive (negative) $m_j$ electron.
}
\label{tab:GS_wo_CH}
\end{table*}

\begin{table*}[]
 \centering
\begin{tabular}{ccc}
 \hline\hline
 $N$ & $J_z$ & Ground State\\
 \hline
 \multirow{2}{*}{2}
 & +2 & $\alpha_2\LocalState{0}{0}{1}{0}{1}{0} - \beta_2\LocalState{1}{0}{0}{0}{1}{0}$ \\
 & -2 & $\alpha_2\LocalState{0}{0}{0}{1}{0}{1} + \beta_2\LocalState{0}{1}{0}{0}{0}{1}$ \\
 \hline
 \multirow{2}{*}{3}
 & +3/2 & $\alpha_3\LocalState{0}{1}{1}{0}{1}{0} - \beta_3\LocalState{0}{0}{1}{1}{1}{0} - \gamma_3\LocalState{1}{1}{0}{0}{1}{0} + \delta_3\LocalState{1}{0}{0}{1}{1}{0}$ \\
 & -3/2 & $\alpha_3\LocalState{1}{0}{0}{1}{0}{1} + \beta_3\LocalState{0}{0}{1}{1}{0}{1} + \gamma_3\LocalState{1}{1}{0}{0}{0}{1} + \delta_3\LocalState{0}{1}{1}{0}{0}{1}$ \\
 \hline
 4 & 0 & $\alpha_4'\LocalState{0}{0}{1}{1}{1}{1} + \gamma_4'\LocalState{1}{1}{1}{1}{0}{0} + \beta_4'\LocalState{1}{1}{0}{0}{1}{1} - \delta_4'\left(\LocalState{1}{0}{0}{1}{1}{1} - \LocalState{0}{1}{1}{0}{1}{1}\right)$\\
 \hline\hline
\end{tabular}
\caption{
	Ground state for a given sector of the local Hamiltonian \emph{without the four-scattering terms}.
	In our notation, the upper (lower) level represents the $j=1/2$ ($3/2$) flavors and the lower left (right) level corresponds to the $m_j=\pm 1/2$ ($m_j=\pm 3/2$) flavor.
	Solid (open) circles mark the positive (negative) $m_j$ electron.
}
\label{tab:GS_wo_FS}
\end{table*}

\end{document}